\documentclass{appolb}
\usepackage[utf8x]{inputenc}
\usepackage[T1]{fontenc}
\usepackage{amstext}
\usepackage{amsfonts}
\usepackage{amsmath}
\usepackage{graphicx}
\usepackage[extendedchars]{grffile}
\usepackage{setspace}
\usepackage{braket}
\usepackage{gensymb}

\newcommand{\abs}[1]{\left|#1\right|}

\newcommand{\ee}{\mathrm e}
\newcommand{\dd}{\mathrm d}

\newcommand{\bee}{\begin{equation}}
\newcommand{\eee}{\end{equation}}
\begin{document}



\title{
\begin{flushright}
IFJPAN-IV-2013-8 \\
\end{flushright}
Numerical solution of the integral form of the resummed Balitsky-Kovchegov equation%
}
\author{K.\ Kutak$^a$, W.\ P{\l}aczek$^b$, D.\ Toton$^a$
\address{$^a${\it H.\ Niewodnicza\'nski Institute of Nuclear Physics, Polish Academy of Sciences\\
ul.\ Radzikowskiego 152, 31-342 Krak\'ow, Poland;\\
$^b$Marian Smoluchowski Institute of Physics, Jagiellonian University,\\
ul.\ Reymonta 4, 30-059 Krak\'ow, Poland.}
}
}
\maketitle
\begin{abstract}
The Balitsky--Kovchegov (BK) evolution equation in its resummed integral form as obtained in \cite{Kutak:2011fu,Kutak:2012yr} is
considered.
We solve it numerically and compare to the unresummed BK equation formulated as an integral equation and to the solution obtained by the {\sf BKsolver} package.
Sensitivity of the solution to an introduced resolution parameter and initial conditions is investigated.
\end{abstract}

\section{Introduction}
Quantum Chromodynamics (QCD) is a theory which
is used to set up the initial conditions for the collisions at the Large Hadron Collider (LHC) as well as to calculate properties of hadronic observables.
At high energies as available at the LHC one enters into a region of the phase space where both the energy and
momentum transfers are high and partons eventually form a dense system which is expected to saturate \cite{Gribov:1984tu,Albacete:2010pg,Dumitru:2010iy,Kutak:2012rf,Dusling:2012cg}.
Indeed, there is gowing evidence that the saturation really takes place
\cite{Albacete:2010pg,Dumitru:2010iy,Kutak:2012rf,Dusling:2012cg}.
The basic perturbative QCD equation which sums up the terms proportional to $\alpha_s^n \ln^m (s/s_0)$ and also accounts for formation of the dense sytem of partons is the Balitsky--Kovchegov (BK) equation \cite{Balitsky:1995ub,Kovchegov:1999yj}.
The BK equation, valid in asymptotic regime, does not take into account coherence effects in emission of gluons.
This property manifests itself as independence on the hard scale associated with the external hard probe.
Recently a framework has been provided in \cite{Kutak:2011fu,Kutak:2012yr,Kutak:2012qk,vanHameren:2012uj,vanHameren:2012if} where both the
dense systems and the hard processes at high energies can be studied. This framework was based on the observation that the BK equation can be rewritten in an exclusive form and further extended to include the coherence effects.
In the study presented here, which is a step towards understanding properties of the equations obtained in  \cite{Kutak:2011fu,Kutak:2012yr,Kutak:2012qk}, we perform a numerical analysis of a new form of the BK equation and compare it to the original formulation (see also \cite{Deak:2012mx}). In particular, we study the dependence of the resummed BK equation on a new scale which has the meaning of a resolution scale. We show that the resummed BK equation agrees with the original one when the resolution scale $\mu$ is already of the  order of $10^{-3}\,$GeV. We also investigate the dependence of the solution on the form of the initial conditions.
\begin{figure}[htb]
\centerline{\includegraphics[width=5cm]{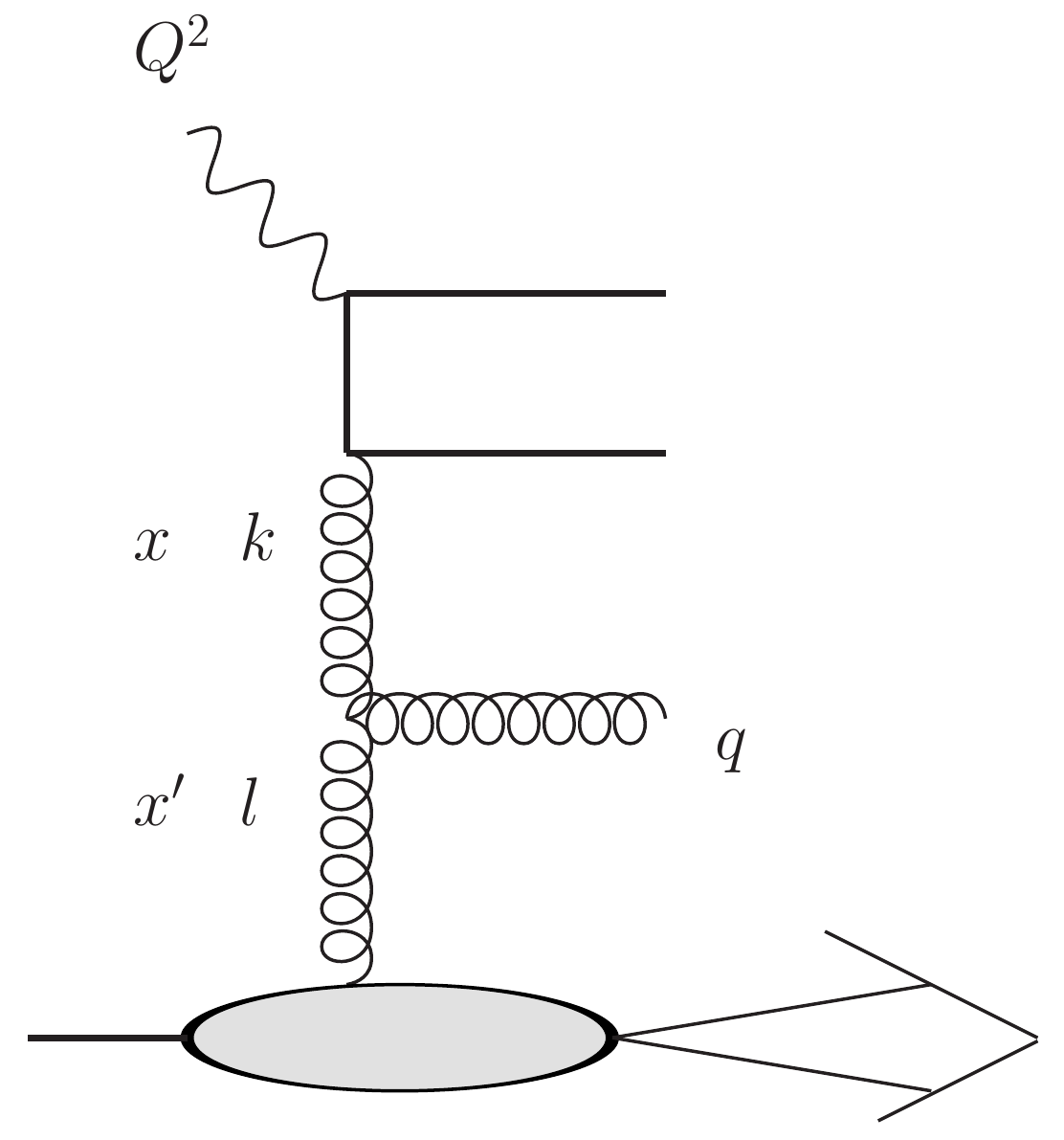}}
\caption{{The diagram explaining the meaning of the variables in the BK equation.}}
\label{fig:Kinematics}
\end{figure}
\section{Exclusive form of the BK equation}
At the leading order in $\ln(1/x)$ the BK equation \cite{Balitsky:1995ub,Kovchegov:1999yj}  for the gluon number density in the momentum space is written as an integral equation and reads \cite{Kutak:2011fu,Kutak:2012yr,Kutak:2012qk}:
\begin{align}
\label{eq:BK}
\Phi(x,k^2)&= \Phi_0(x,k^2)\\\nonumber
&+\overline\alpha_s\int_{\frac{x}{x_0}}^1 \frac{dz}{z}\int_0^{\infty}\frac{dl^2}{l^2}
\bigg[\frac{l^2\Phi(x/z,l^2)- k^2\Phi(x/z,k^2)}{|k^2-l^2|}+ \frac{
k^2\Phi(x/z,k^2)}{\sqrt{(4l^4+k^4)}}\bigg]\\\nonumber
&-\frac{\overline\alpha_s}{\pi R^2}\int_{\frac{x}{x_0}}^1 \frac{dz}{z}\Phi^2(x/z,k^2),
\nonumber
\end{align}
where the  lengths of transverse vectors lying in the transversal plane to the collision axis are $k\equiv|{\bold k}|$, $l\equiv|{\bold l}|$ (${\bold k}$ is a
vector sum of transversal momenta of emitted gluons during evolution), $z=x/x' $(see Fig.~\ref{fig:Kinematics}), $\overline\alpha_s=N_c\alpha_s/\pi$. The strength of the nonlinear term is controlled by the targets radius $R$.
The linear term in eq. (\ref{eq:BK}) can be linked to the process of creation of gluons while the nonlinear term
can be linked to fusion of gluons. The interplay of these two terms as a net effect leads to saturation of gluons.
The unintegrated gluon density obeying the high energy factorization theorem \cite{Catani:1990eg} is obtained from
 \cite{Braun:2000wr,Kutak:2003bd}:
\bee
{\cal F}_{BK}(x,k^2)=\frac{N_c}{\alpha_s \pi^2}k^2\nabla_k^2 \Phi(x,k^2) ,
\label{eq:glue}
\eee
where the angle-independent Laplace operator is given by
$\nabla_k^2=4\frac{\partial}{\partial k^2}k^2\frac{\partial}{\partial k^2}$.\\
As shown in \cite{Kutak:2011fu,Kutak:2012yr} this equation can be rewritten in a resummed form:
\bee
\label{eq:resBK}
\Phi(x,k^2)=\tilde \Phi^0(x,k^2)
+\overline\alpha_s\int_{\frac{x}{x_0}}^1d\,z\int\frac{d^2{\bf q}}{\pi q^2}\,
\theta(q^2-\mu^2)\frac{\Delta_R(z,k,\mu)}{z}\Bigg[\Phi(\frac{x}{z},|{\bf k} +{\bf q}|^2)\nonumber
\eee
\bee
-\frac{1}{\pi R^2}q^2\delta(q^2-k^2)\,\Phi^2(\frac{x}{z},q^2)\Bigg].
\eee
where ${\bf q}={\bf l}-{\bf k}$ and $\Delta_R(z,k,\mu)\equiv\exp\left(-\overline\alpha_s\ln\frac{1}{z}\ln\frac{k^2}{\mu^2}\right)$ is the Regge form factor.

Eq. (\ref{eq:resBK}) is a form of the BK equation in which the resummed terms resulting in
the Regge form factor are the same for the linear and nonlinear part.
This form served as a guiding equation to generalize the CCFM  equation \cite{Ciafaloni:1987ur,Catani:1989sg,Catani:1989yc} to the KGBJS   equation \cite{Kutak:2011fu,Kutak:2012yr} which includes nonlinear effects. These effects allow for recombination of partons with a constraint on an emission angle.
\section{Computational method}
In order to study the behaviour of both the equations (\ref{eq:BK}) and (\ref{eq:resBK}),
they were solved numerically following an iterative procedure,
which we detail in this section.\\
For numerical treatment of Eq. (\ref{eq:BK}),
the inner integral is approximated
 by reducing it to a finite interval $\left(k_0, q_f\right)$.
This allows for a direct numerical representation of the $\Phi$ function,
which is the solution to be computed.
Iterative refinement of it
is more naturally expressed with
$w = \frac x z$ as the integration variable.
Also, to give an accurate view of the numerical procedure,
we explicitely denote the second argument of $\Phi$ as $k_2 = k^2$, $l_2=l^2$,
so the equation reads
\begin{align}
\Phi(x, k_2)
&=
\Phi_0(x, k_2)
\nonumber \\
&+\bar \alpha_s
\int_x^{x_0}
\frac {\dd w} w
\int_{k_0^2}^{q_f^2}
\frac{\dd l_2}{l_2}
\left[
\frac{l_2 \Phi(w, l_2) - k_2 \Phi(w, k_2)}
{\abs{k_2 - l_2}}
+
\frac{k_2 \Phi(w, k_2)}
{\sqrt{4 l_2^2 +  k_2^2}}
\right]
\nonumber \\ &-
\frac
{\bar\alpha_s}
{\pi R^2}
\int_x^{x_0}
\frac {\dd w} w\,
\Phi^2 \left(w, k_2\right),
\end{align}
where for simplicity we assume
$R = \frac {1} {\sqrt  {\pi}}$  and
$\bar \alpha_s = \frac {N_c \alpha_s} {\pi}$ with the fixed QCD coupling constant $\alpha_s = 0.2$.
The evolution starts at $x_0 = 10^{-2}$.\\
The function $\Phi$ is represented on a regular mesh of points $(x_m, {k_2}_n)$
that are equidistant in a logarithmic setting: $\log x_m = \log x_{min} + m \Delta_x$ (and simlilarly in $k_2$).
For values of $\Phi$ with arbitrary arguments
we employ a bilinear interpolation with respect to logarithms of the variables.
The iterative procedure starts with $\Phi_0$ as a first approximation of the solution.
Then the right hand side of the equation is calculated repeatedly to provide and use subsequent approximations.
Since Eq.  (\ref{eq:BK}) can be cast as a differential equation in $x$ as well,
the above algorithm can be modified to be more efficient.
Namely, the particular structure of the equation
permits us to work with the domain of $\Phi$ reduced in $x$
to cover the arbitrarily small interval $(x', x_0)$.
In order to extend some known solution to reach some smaller $x'$,
it is enough that the iterative recalculation of the right-hand side
is performed only for the newly considered values of $x$.
Starting with $x'=x_0$ and going down to next grid points we save computational time.
In our implementation,
the integrals are computed using the {\sf VEGAS} method, as available in the {\sf CUBA} library \cite{Hahn:2004fe}.
To ease the integration,
extra variable changes:
$\dd w/w \rightarrow \dd \left(\ln w\right)$
 and
$\dd l_2/l_2 \rightarrow \dd \left(\ln l_2\right)$
are employed.

The following two forms of the driving term $\Phi_0$ are considered here:
\begin{align}
\Phi^a_0\left(x, k_2\right) &= \exp\left(-k_2/\text{GeV}^2\right),\\
\Phi^b_0\left(x, k_2\right) &= \left(k_2/\text{GeV}^2\right)^{-1/2}.
\end{align}

The resummed form of the BK equation, eq.~(\ref{eq:resBK}), is solved in approximated form for similar reasons.
While the lower limit of the inner integral is now exactly $\mu$,
the upper limit and the vector sum $\abs{{\mathbf{k}} + {\mathbf{q}}}= S(q_2, k_2, \varphi)$ need to be constrained:
\begin{align}
\Phi(x, k_2)
&=
\tilde \Phi_0(x, k_2)
\nonumber \\
&+
\bar \alpha_s
\int_x^{x_0}
\frac {\dd w} w
\Delta_R(z, k_2)
\int_{\mu^2}^{q_f^2}
\frac {\dd q_2} {\pi q_2}
\int_{0}^{\pi}
\dd \varphi\,
\Phi\left(w, m (S(q_2, k_2, \varphi), k_0^2, q_f^2)\right)
\nonumber \\
&-
\frac {\bar \alpha_s}
{\pi R^2}
\int_x^{x_0}
\frac {\dd w} w
\Delta_R(z, k_2)
\Phi^2(w, k_2).
\end{align}

The length of the vector sum
\begin{equation}
S(a_2, b_2, \alpha) = a_2 + b_2 + 2 \sqrt{a_2 b_2} \cos \alpha
\end{equation}
may fall outside the $\left({k_0^2},{q_f^2}\right)$ range,
hence the integrand is approximated by limiting the second argument of $\Phi$.
This is done as follows:
\begin{equation}
m\left(k'^2, k_0^2, q_f^2\right) =
\begin{cases}
k_0^2 &\text{ when } k'^2 < k_0^2, \\
k'^2 & \text{ when }  k_0^2 \leq k'^2 \leq q_f^2,\\
q_f^2 &\text{ when } k'^2 > q_f^2.
\end{cases}
\end{equation}
This way the domains of $\Phi$ and of the integrals all can be kept finite.

The initial conditions resulting from the resummation procedure include an extra factor as follows:
\begin{equation}
\tilde\Phi^{a,b}_0(x, k_2) =
\ee^{
- \bar \alpha_s
\ln \frac {x_0} x
\ln \frac {k_2}{\mu^2}
}
\Phi^{a,b}_0(x, k_2),
\end{equation}

Except for the above remarks, both equations are solved using the same method.
The numerical solutions presented below are computed
with $k_0 = 0.001 \text{ GeV}$ and $q_f = 100 \text{ GeV}$, for
the cut-off values  ranging from $\mu = k_0$ to $\mu = 0.01 \text{ GeV}$.


\section{Numerical solutions}
\begin{figure}[t!]
\centerline{
 \includegraphics[width=7cm,trim=5.5cm 1cm 3.5cm 1cm]{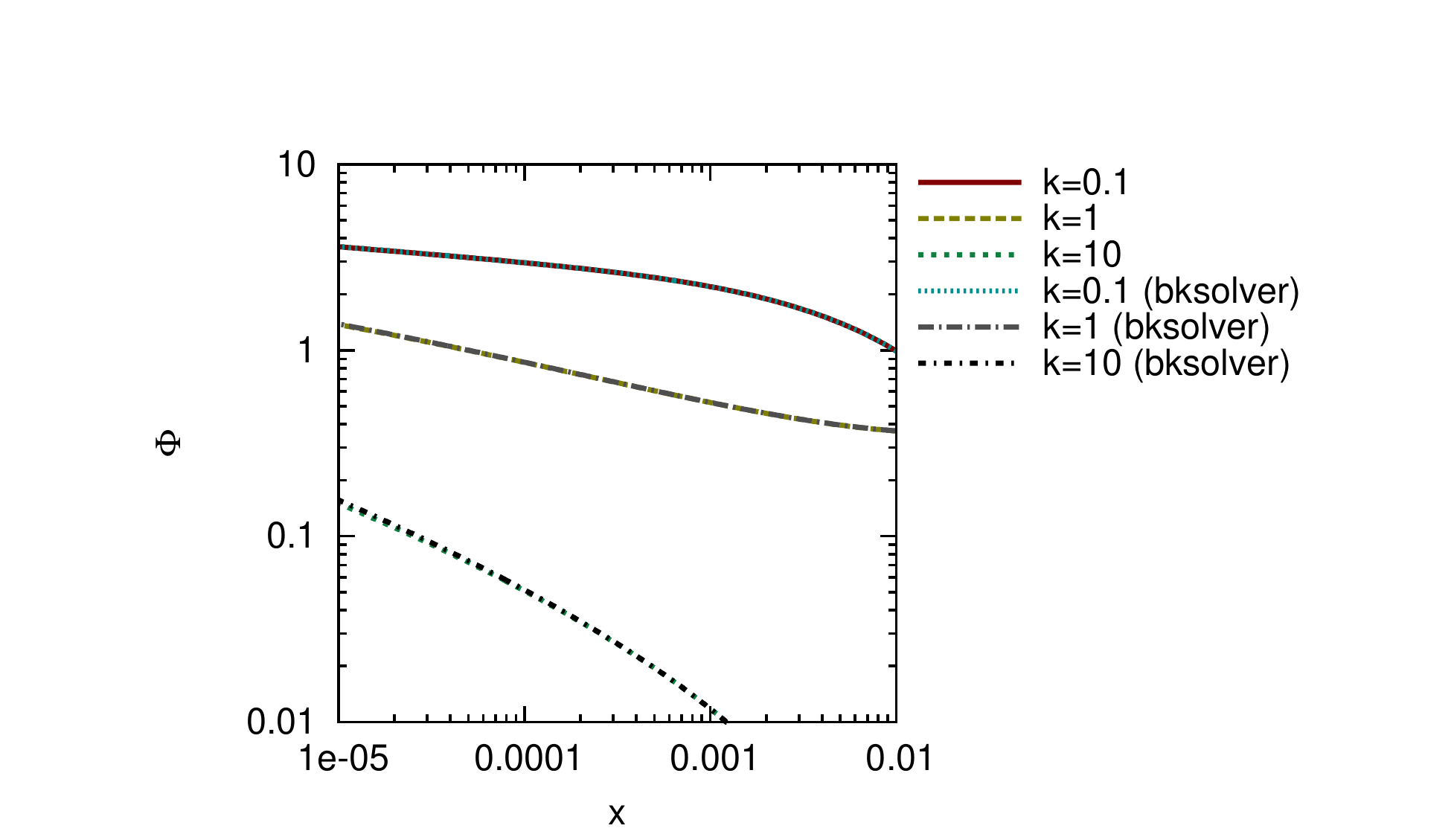}
}\centerline{
 \includegraphics[width=7cm,trim=5.5cm 1cm 3.5cm 1cm]{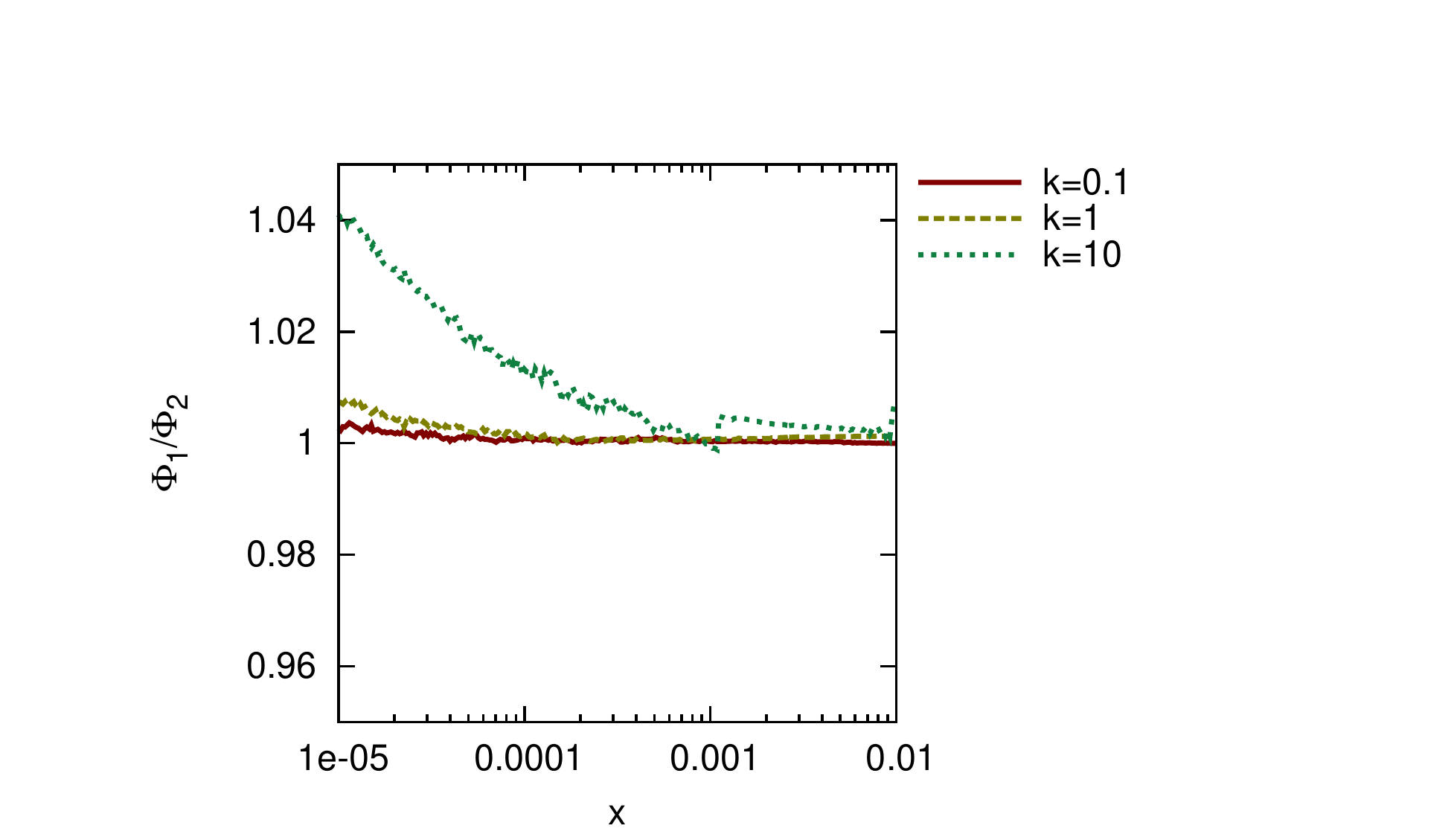}
}
\caption{Upper plot: the solutions of the BK equation as function of $x$ as formulated in eq.~(\ref{eq:BK}) compared with the solutions obtained  with the {\sf BKsolver}  package \cite{Enberg:2005cb}.
Lower plot: ratios of these solutions.
}
\label{fig:plot2a}
\end{figure}

\begin{figure}[htb]
\centerline{
 \includegraphics[width=7cm,trim=5.5cm 0cm 3.5cm 1cm]{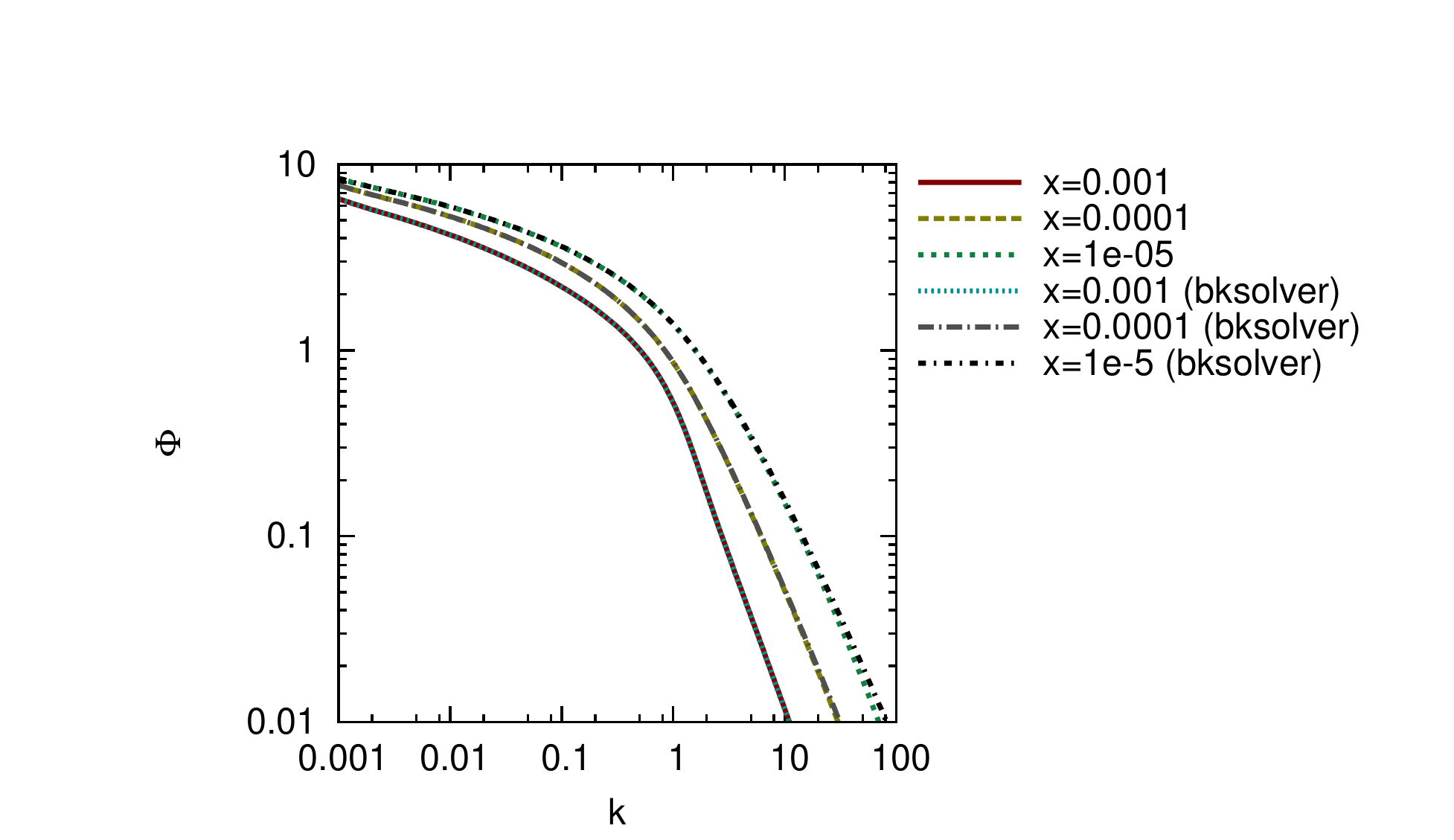}
}\centerline{
 \includegraphics[width=7cm,trim=5.5cm 0cm 3.5cm 1cm]{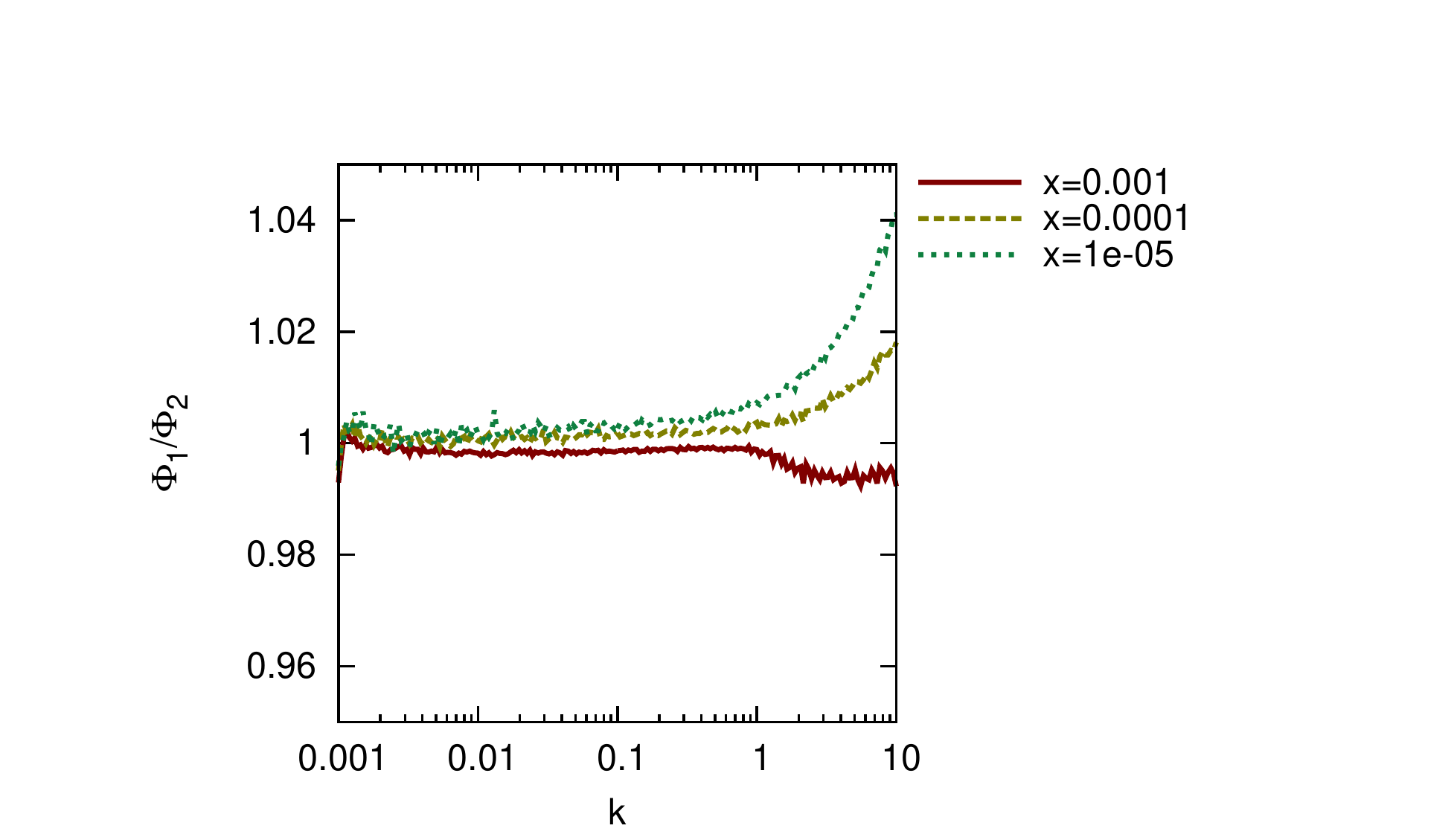}
}
\caption{Upper plot: the solutions of the BK equation as function of $k$ as formulated in eq.~(\ref{eq:BK}) compared with the solutions obtained  with the {\sf BKsolver}  package \cite{Enberg:2005cb}.
Lower plot: ratios of these solutions.
}
\label{fig:plot2b}
\end{figure}

\begin{figure}[t!]
\centerline{
 \includegraphics[width=7cm,trim=5.5cm 0 3.5cm 1cm]{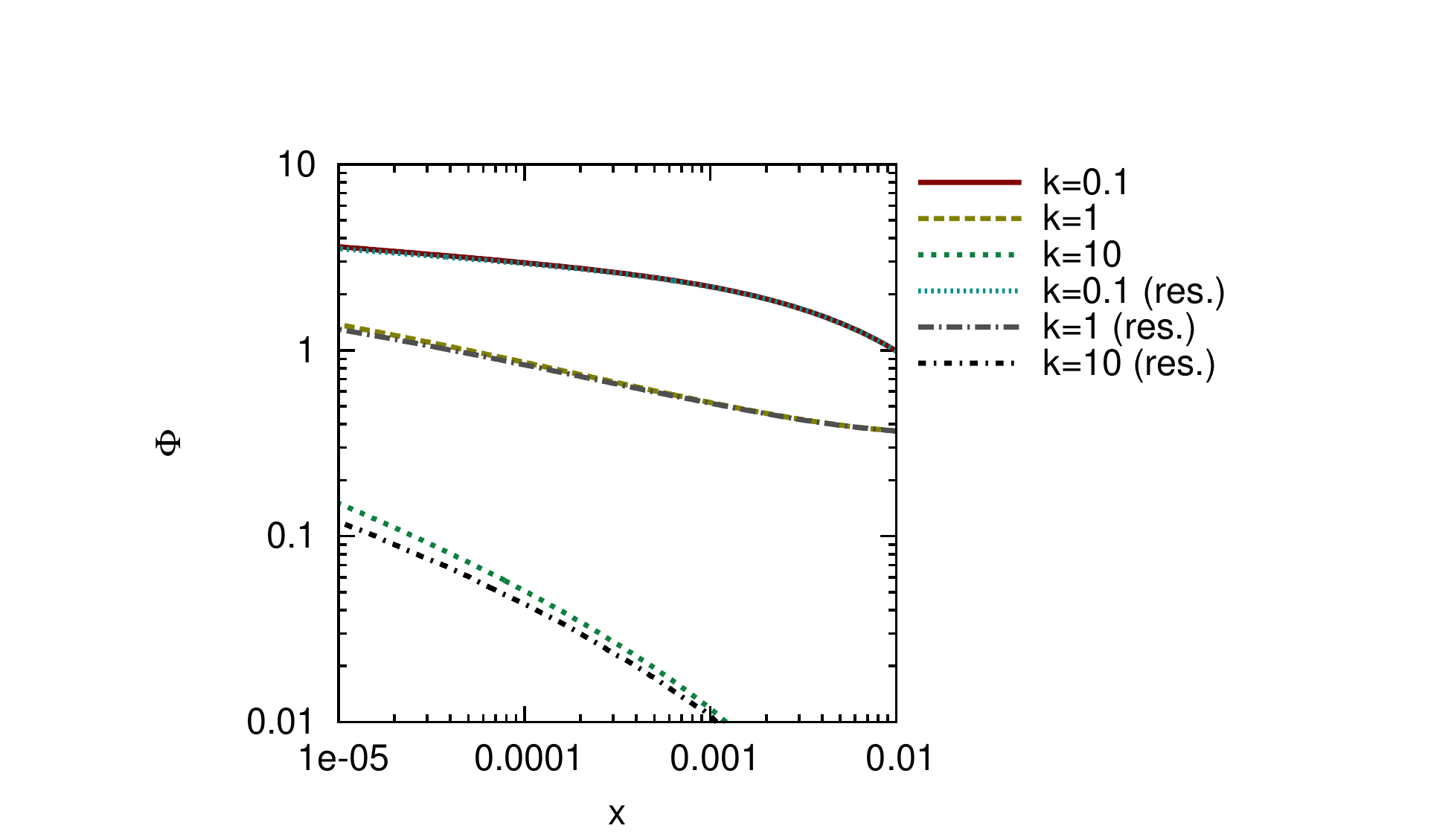}
}\centerline{
 \includegraphics[width=7cm,trim=5.5cm 0 3.5cm 1cm]{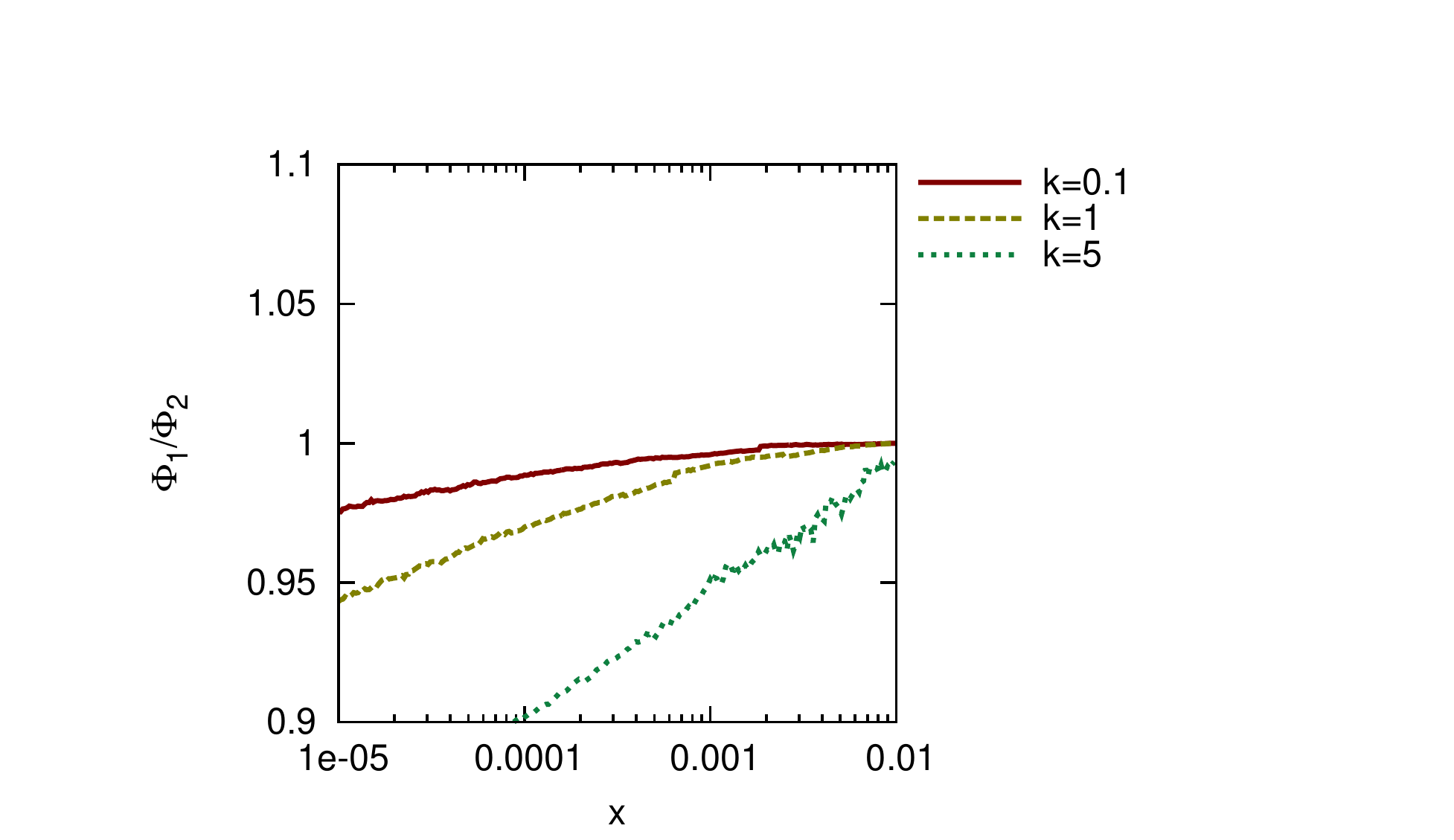}
}
\caption{The comparison of the solutions of the BK equation as formulated in eq.~(\ref{eq:BK}) with the solutions of its resummed form
given in eq.~(\ref{eq:resBK}); $\mu = k_0 = 0.001 \text{ GeV}$.
Lower plot: ratios of these solutions with values obtained for eq.~(\ref{eq:resBK}) in the denominator.
}
\label{fig:plot3a}
\end{figure}

\begin{figure}[htb]
\centerline{
 \includegraphics[width=7cm,trim=5.5cm 0 3.5cm 1cm]{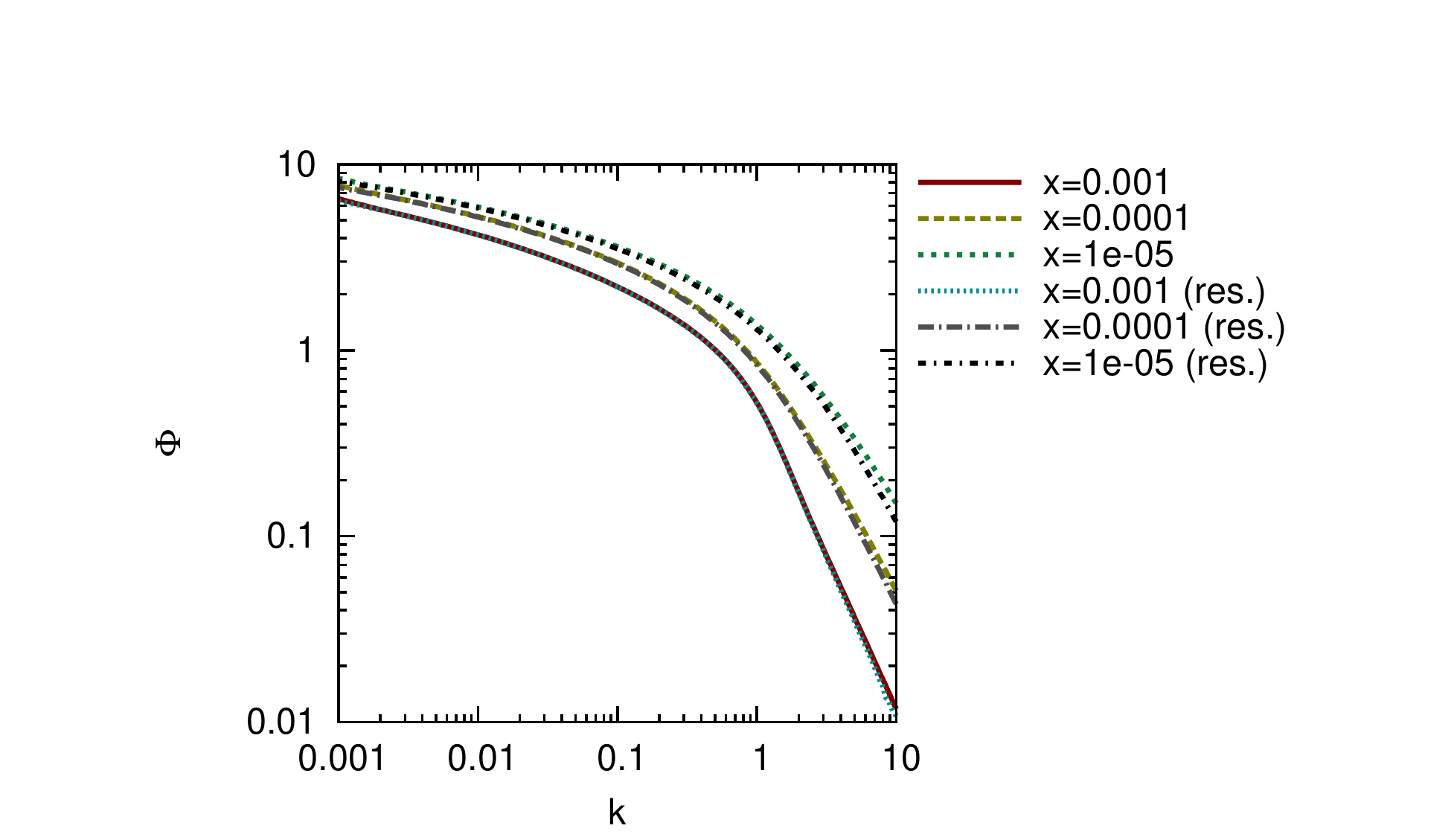}
}\centerline{
 \includegraphics[width=7cm,trim=5.5cm 0 3.5cm 1cm]{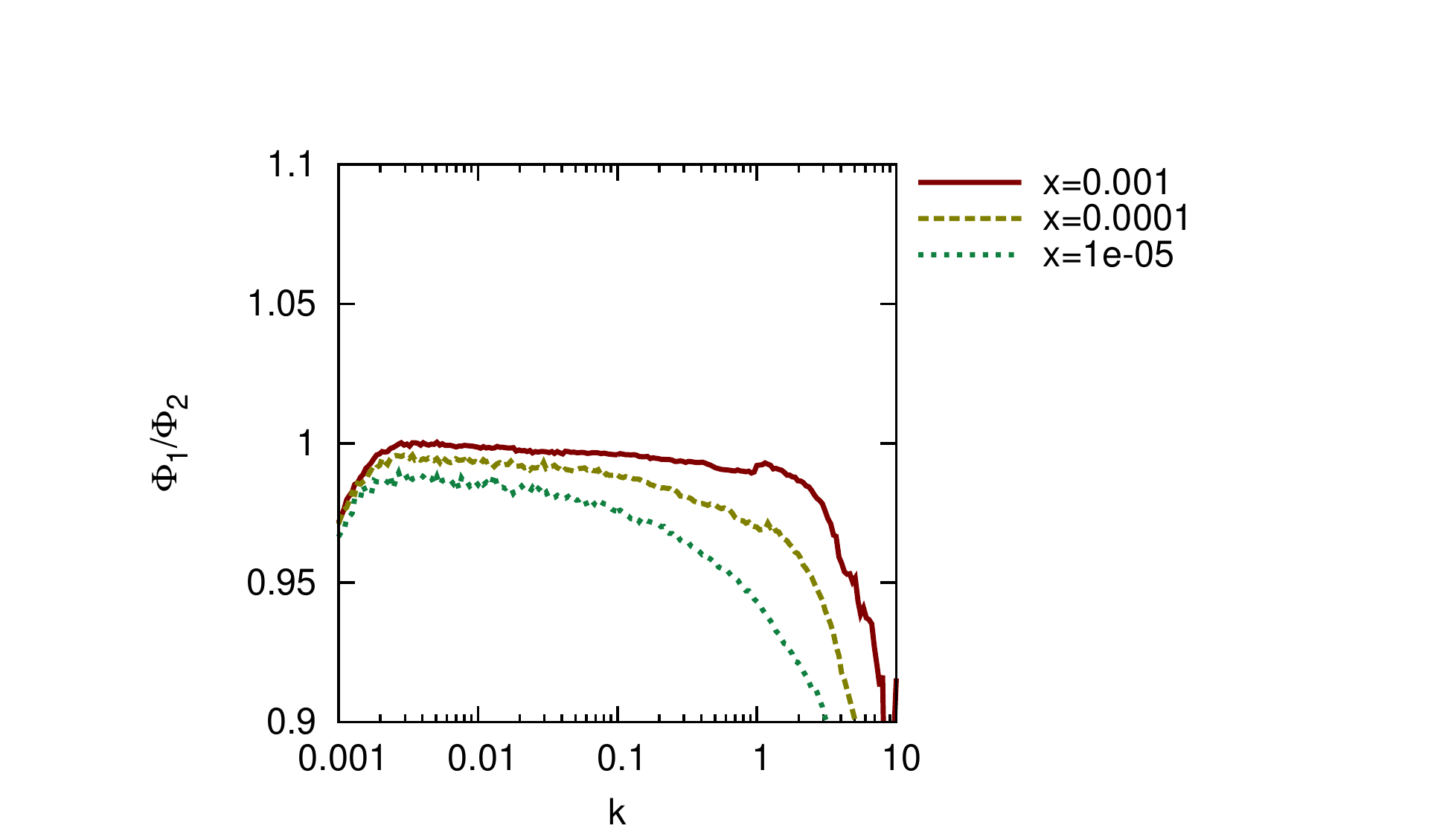}
}
\caption{Cross-section along $k$ of the soltions compared on Fig. \ref{fig:plot3a}.
}
\label{fig:plot3b}
\end{figure}

\begin{figure}[t!]
\centerline{
 \includegraphics[width=7cm,trim=5.5cm 0 3.5cm 1cm]{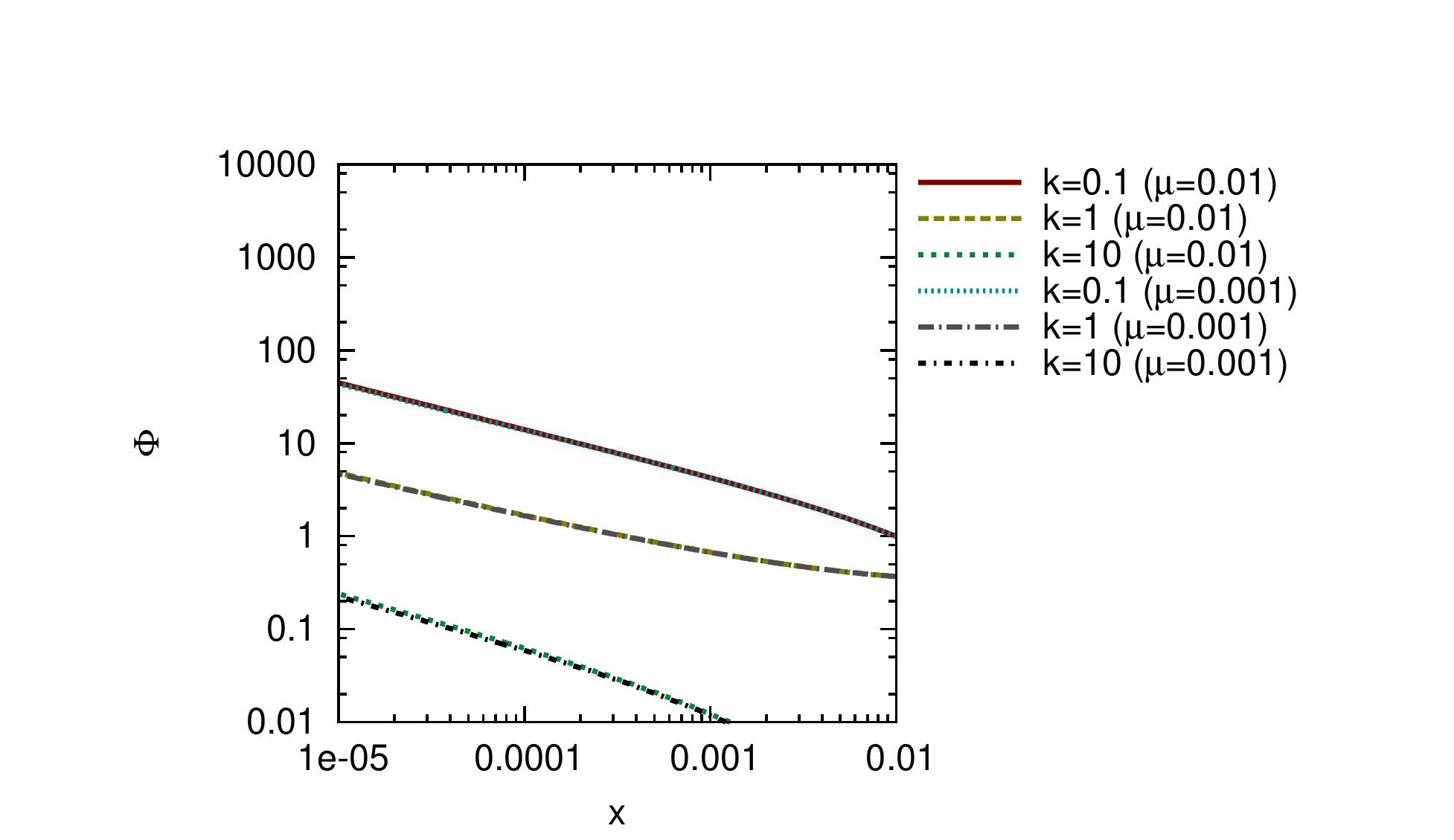}
}\centerline{
 \includegraphics[width=7cm,trim=5.5cm 0 3.5cm 1cm]{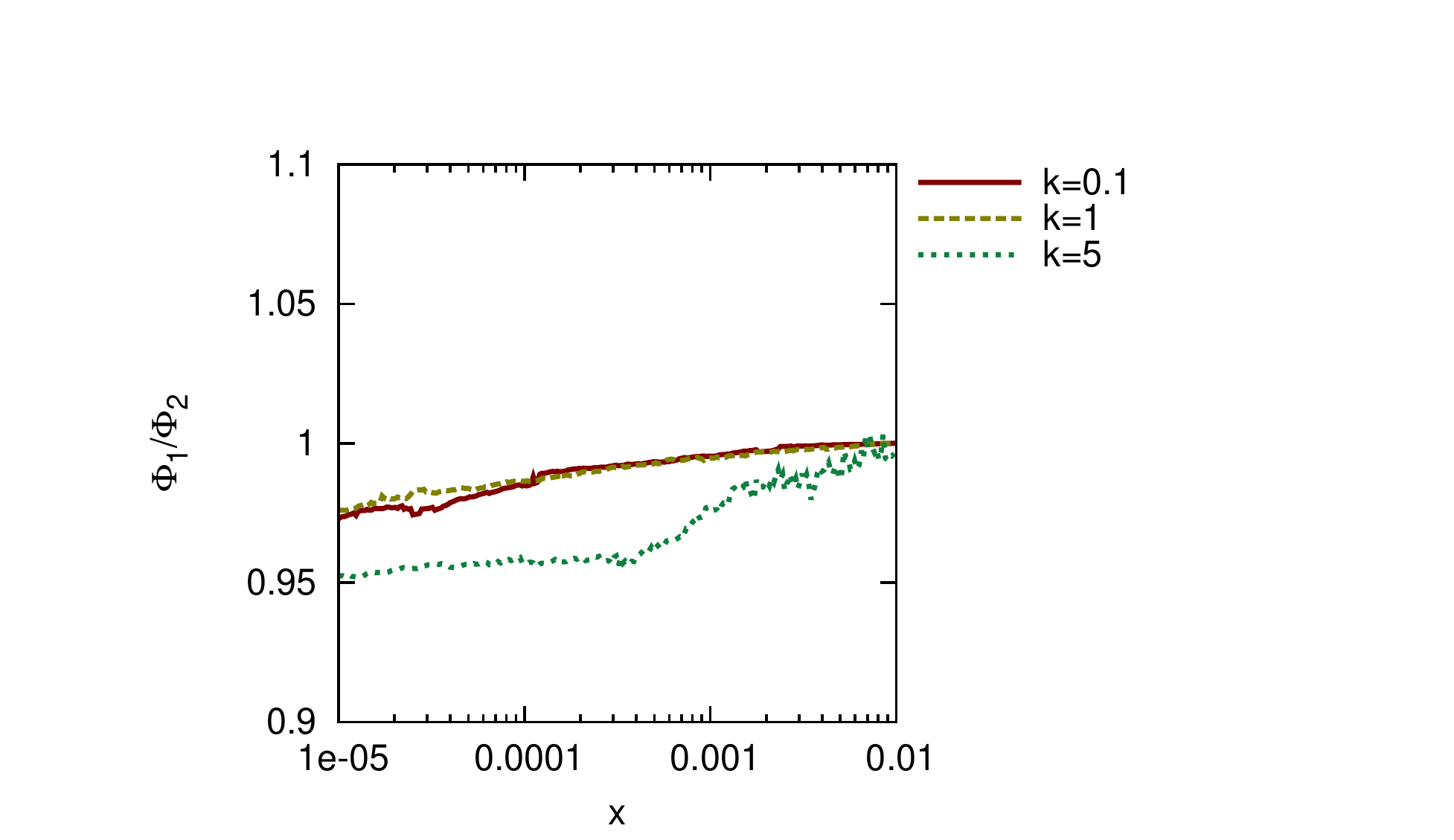}
}
\caption{Sensitivity to the cut-off $\mu$ without the nonlinear term.
Upper plot: solutions of the resummed BFKL equation, i.e. eq.~(\ref{eq:resBK}) with the nonlinear term neglected; their ratios are plotted on the lower pane.}
\label{fig:plot4aa}
\end{figure}

\begin{figure}[htb]
\centerline{
 \includegraphics[width=7cm,trim=5.5cm 0 3.5cm 1cm]{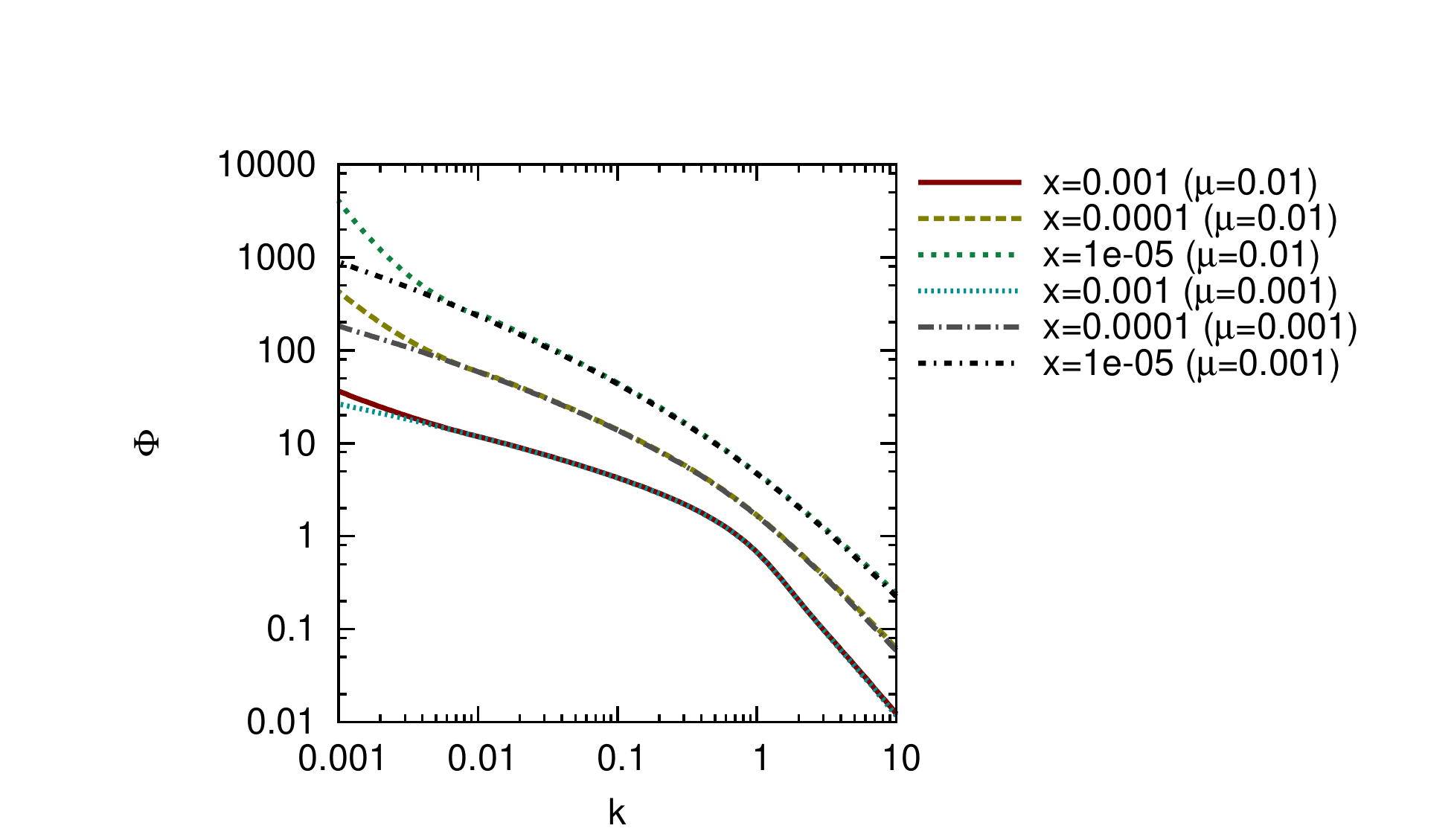}
}\centerline{
 \includegraphics[width=7cm,trim=5.5cm 0 3.5cm 1cm]{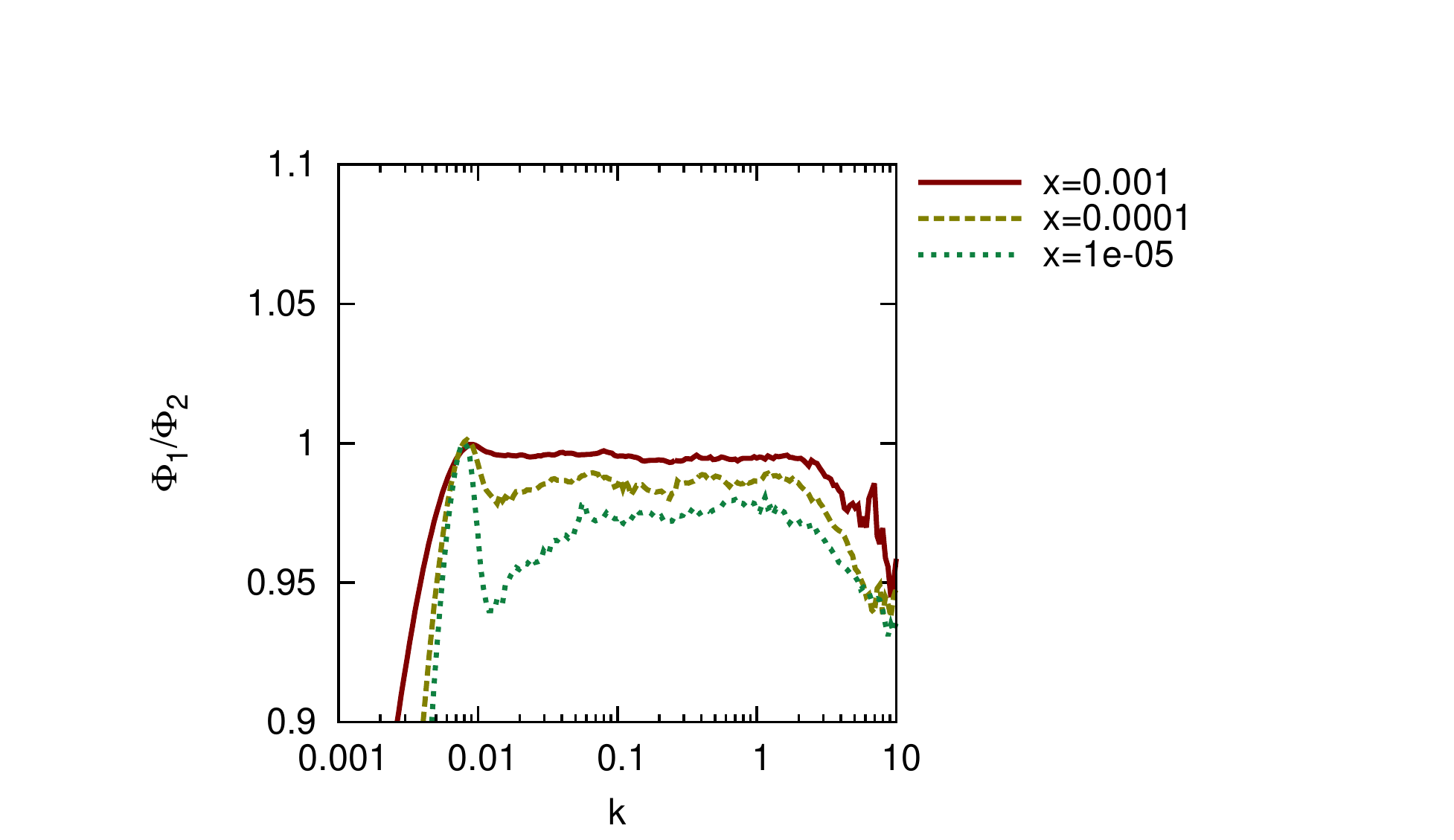}
}
\caption{Cross-section along $k$ of the solutions compared on Fig. \ref{fig:plot4aa}
}
\label{fig:plot4ab}
\end{figure}

\begin{figure}[t!]
\centerline{
 \includegraphics[width=7cm,trim=5.5cm 0 3.5cm 1cm]{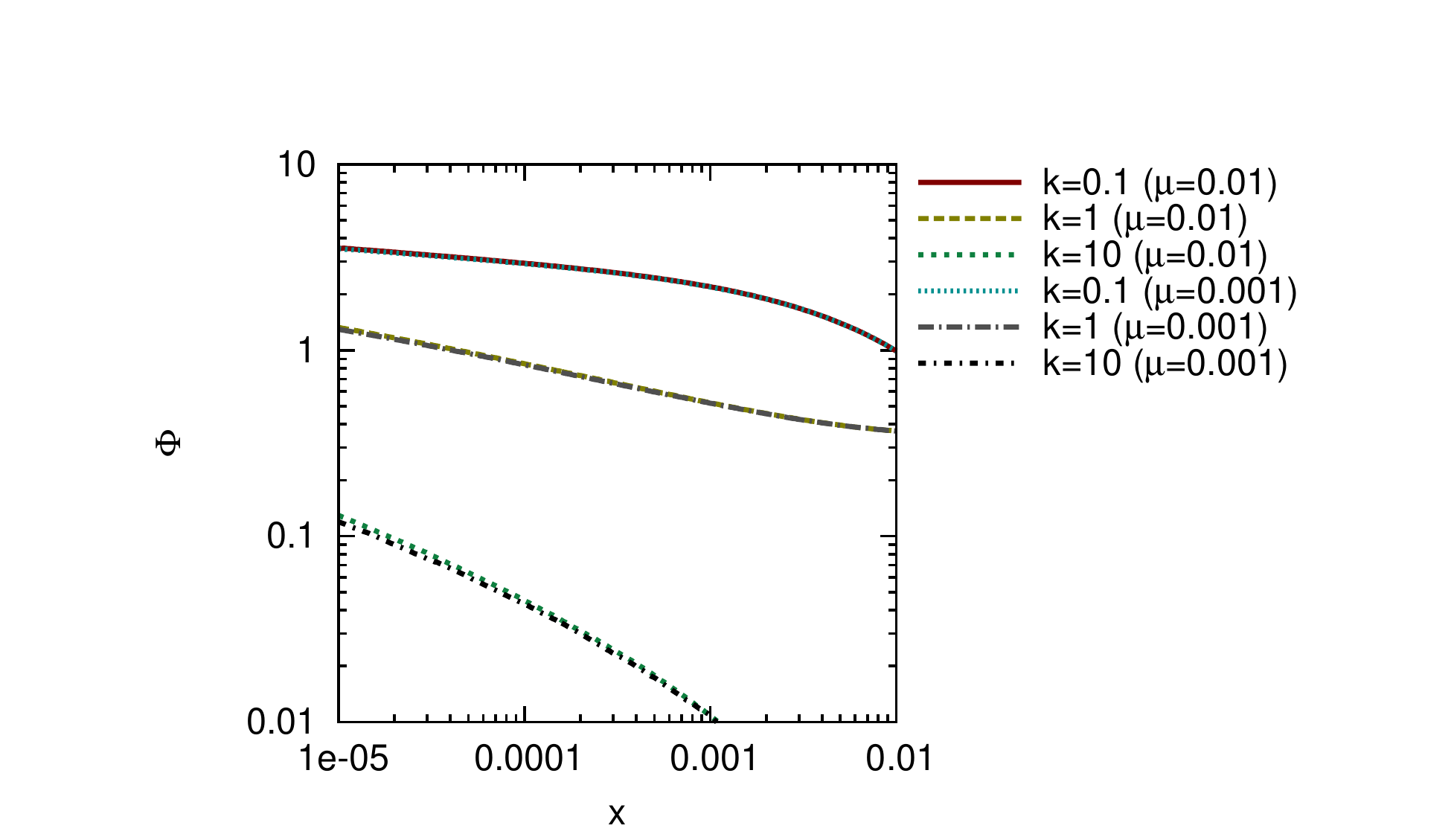}
}\centerline{
 \includegraphics[width=7cm,trim=5.5cm 0 3.5cm 1cm]{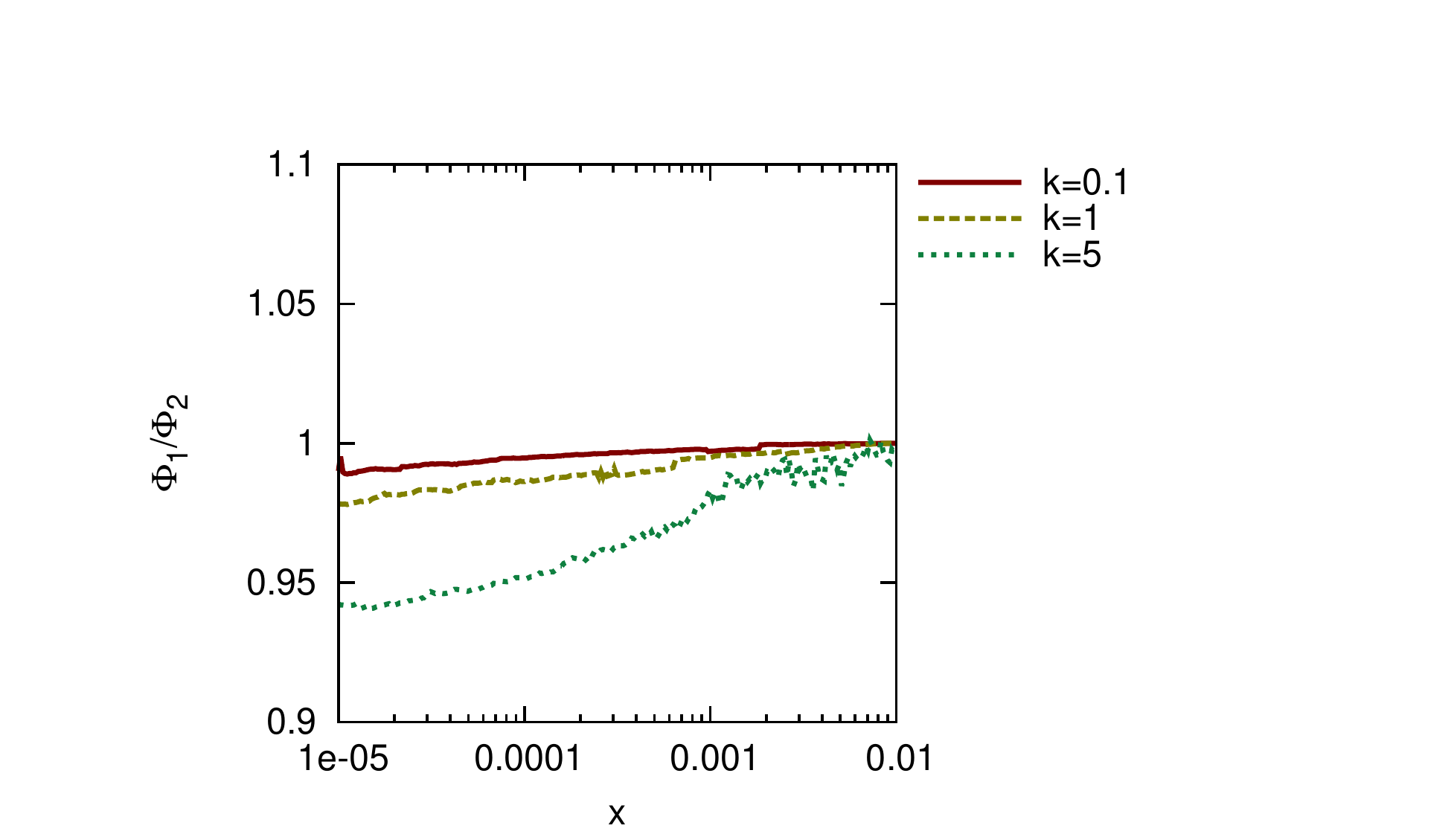}
}
\caption{The effect of the nonlinear term on the sensitivity to the cut-off $\mu$.
The solutions of the resummed BK equation of eq.~(\ref{eq:resBK}); for $\mu = 0.01 \text{ GeV}$ (red and next lines) and $\mu = 0.001 \text{ GeV}$ (blue and next lines).
Lower plot: ratios of these solutions.}
\label{fig:plot4ba}
\end{figure}

\begin{figure}[htb]
\centerline{
 \includegraphics[width=7cm,trim=5.5cm 0 3.5cm 1cm]{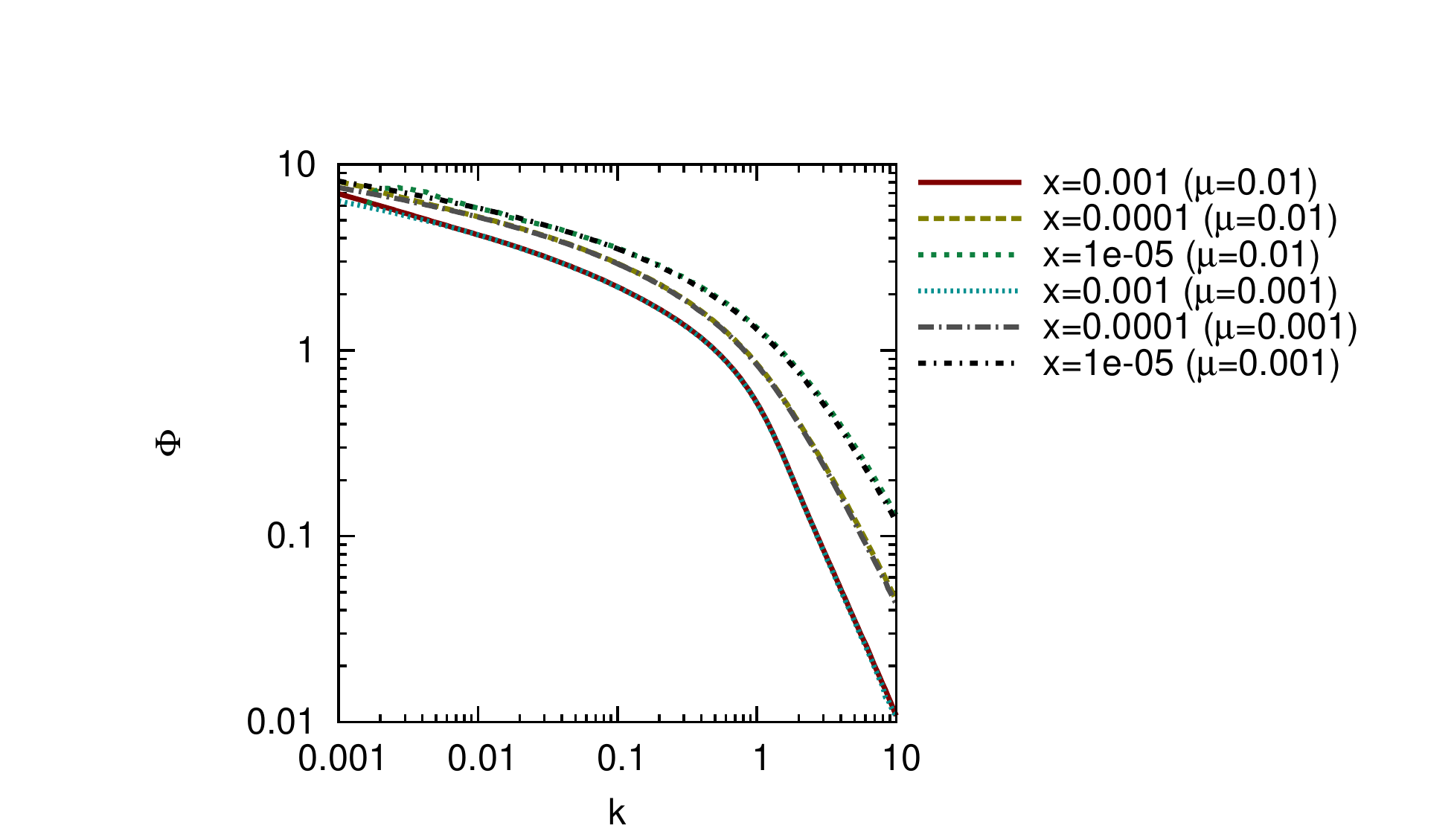}
}\centerline{
 \includegraphics[width=7cm,trim=5.5cm 0 3.5cm 1cm]{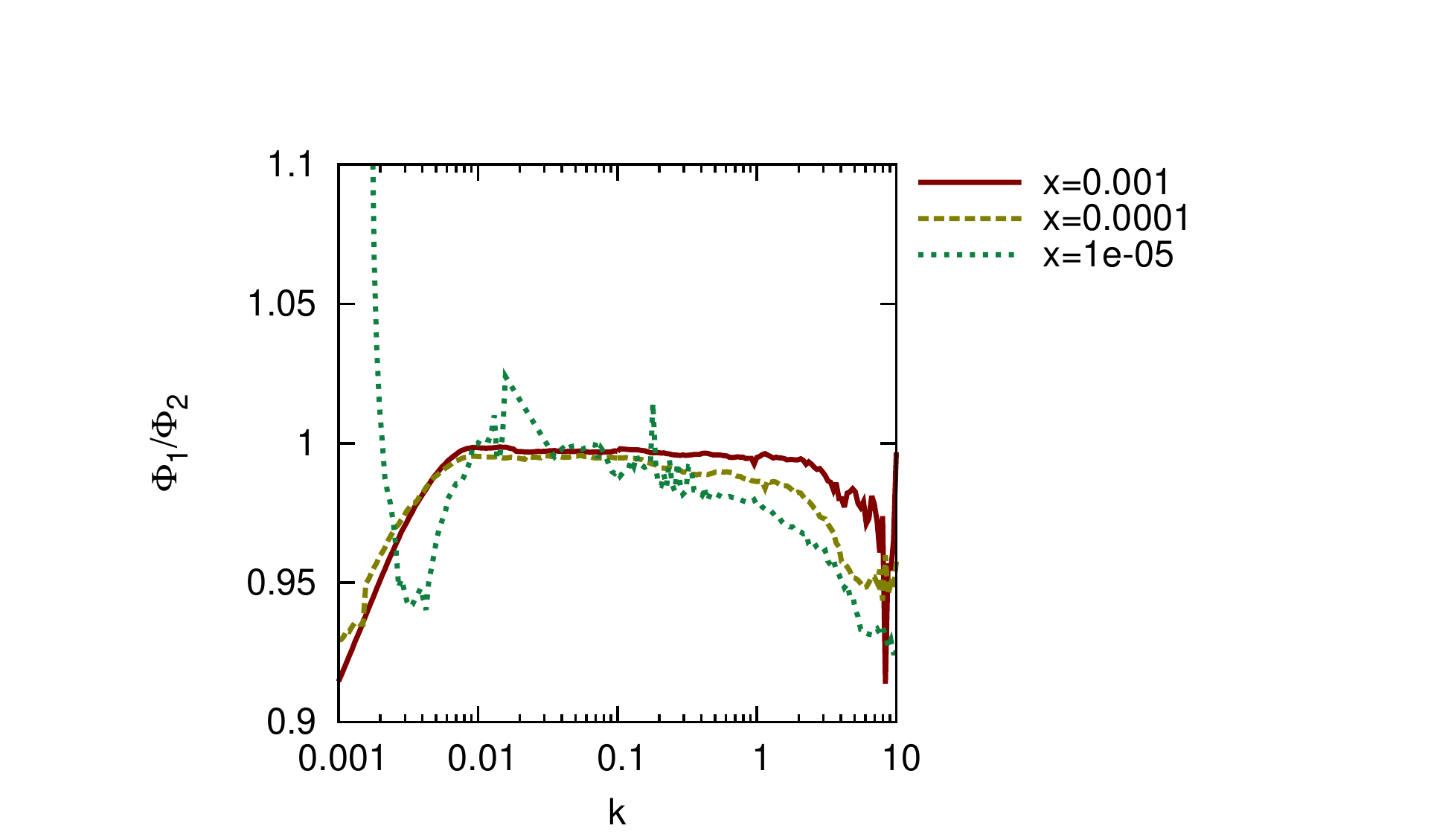}
}
\caption{Cross-section along $k$ of the solutions compared on Fig. \ref{fig:plot4ba}
}
\label{fig:plot4bb}
\end{figure}

\begin{figure}[t!]
\centerline{
 \includegraphics[width=7cm,trim=5.5cm 0 3.5cm 1cm]{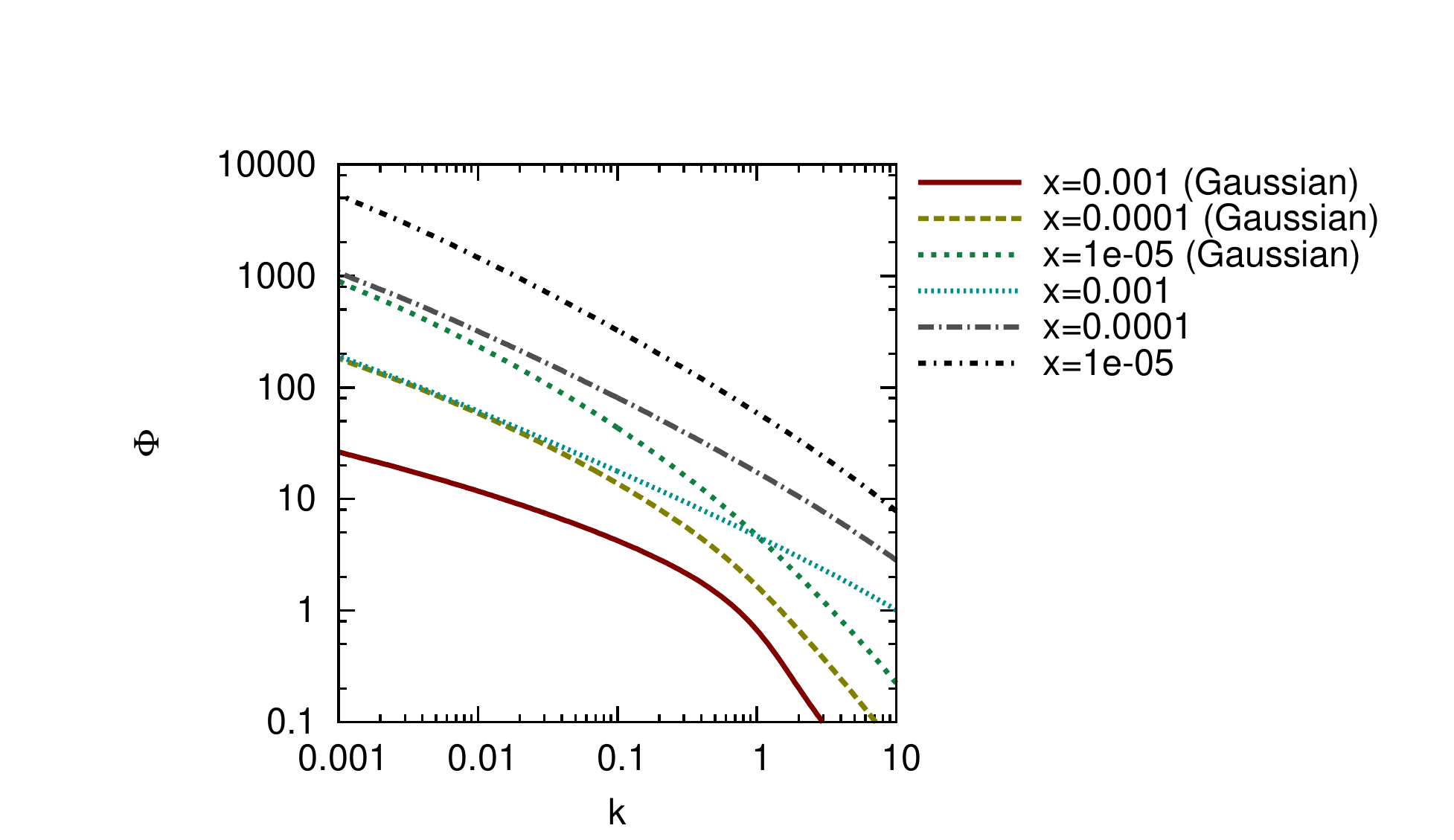}
}\centerline{
 \includegraphics[width=7cm,trim=5.5cm 0 3.5cm 1cm]{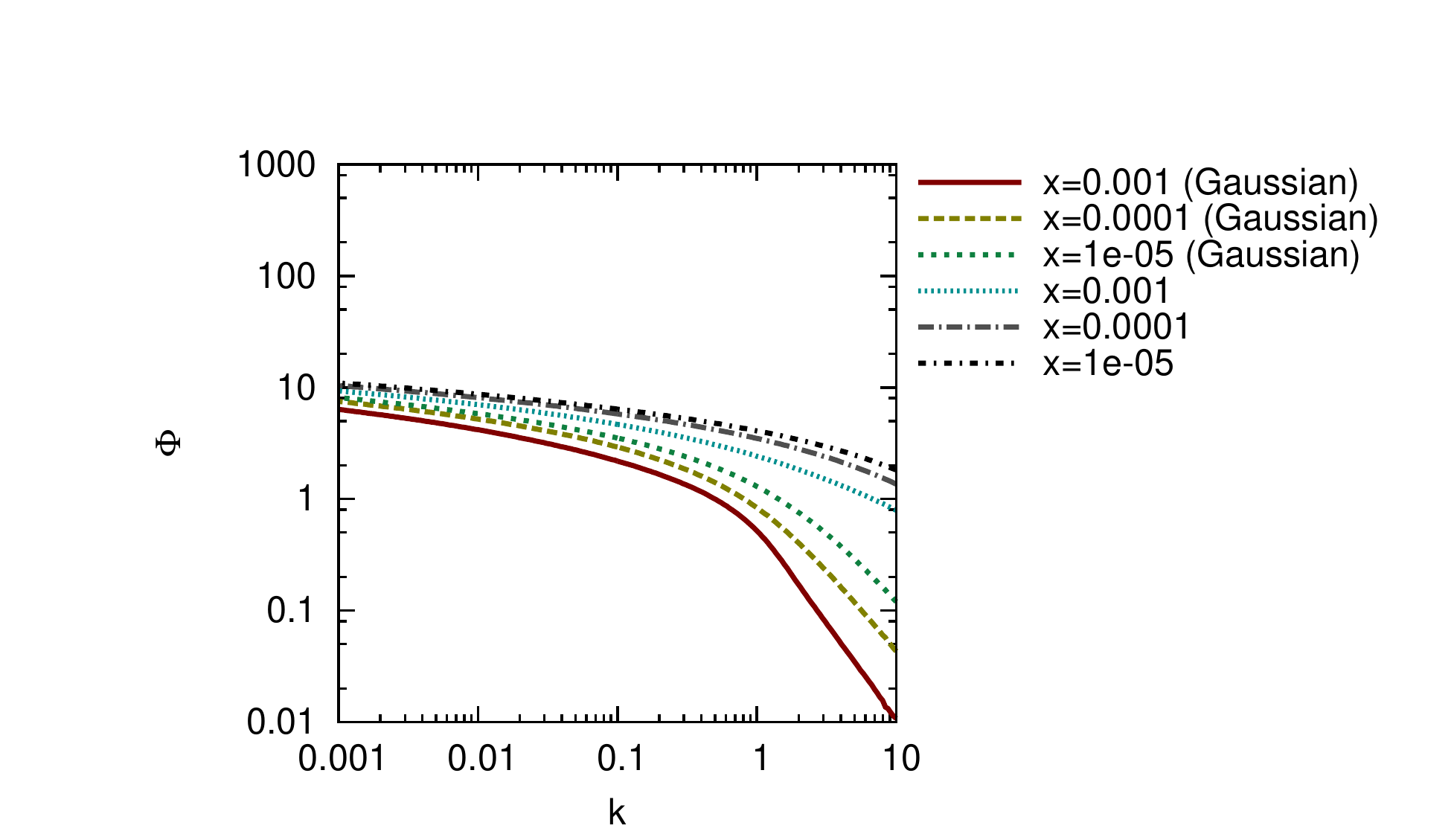}
}
\caption{The comparison of the solutions obtained with different forms of the $\tilde \Phi_0$ term. Left: the resummed BFKL equation, i.e. eq.~(\ref{eq:BK}) with the nonlinear term neglected; right: the resummed BK equation of eq.~(\ref{eq:resBK});
red lines: $\tilde \Phi_0=\ee^{
- \bar \alpha_s
\ln \frac {x_0} x
\ln \frac {k_2}{\mu^2}
}
\exp\left(-k^2/\text{GeV}^2\right)$,
blue lines: $\tilde \Phi_0=
\ee^{
- \bar \alpha_s
\ln \frac {x_0} x
\ln \frac {k_2}{\mu^2}
}
(k^2/\text{GeV}^2)^{-1/2}$.
}
\label{plot5}
\end{figure}

In this section we present results of numerical solution of the considered equations.
In Figs.~\ref{fig:plot2a} and \ref{fig:plot2b} we present the solutions of the BK equation formulated as an integral equation compared to
the more commonly used integro-differential formulation. The former is solved via the iteration method, while the later by using the {\sf BKsolver} package \cite{Enberg:2005cb}.
As can be seen,
the solutions stay within 1\% for $k<1\text{ GeV}$.
Above this value,
the solutions diverge
and at the lowest $x$ considered the relative difference reaches 4\%.
For the largest values of $k$, the distribution $\Phi$ approaches zero anyway,
thus the agreement between the two solutions is satisfactory.
As the next result we present in Figs.~\ref{fig:plot3a} and \ref{fig:plot3b} the solution of the resummed form of the BK equation and compare it to the unresummed one. In the former, the scale $\mu$ introduces numerically some weak dependence of the solution on its value. This dependence is expected formally to disappear in the limit $\mu\rightarrow 0$.
In Figs.~\ref{fig:plot4aa}, \ref{fig:plot4ab}, \ref{fig:plot4ba} and \ref{fig:plot4bb} we study the dependence of the solution of the BFKL and BK equations on the parameter $\mu$. As already mentioned above, the resummation procedure assumes that the scale $\mu$ is the smallest scale in the problem and that it should tend to zero. One can see that the BFKL equation is more sensitive to the resummation parameter than the BK one. The reason for this is the feature of the saturation scale which provides a cut-off on small momenta and therefore weakens the dependence on $\mu$.
Particularly interesting is the shape of the gluon density from the BFKL equation
shown in the upper panel of Fig.~\ref{fig:plot4ab}.
One can see that when the cut-off is larger than the probed $k_t$ value, the distribution bends upwards. This is due to the fact that the Regge form factor in that case becomes larger than one. When we lower the cut-off, this structure disappears.
Finally, in Fig.~\ref{plot5} we investigate the dependence of the solutions on the form of the initial conditions. The comparison of the BFKL and BK cases indicates that the BK equation is more universal, i.e. the spread of its solutions is smaller.
\section{Summary}
In this paper we have performed the study of the resummed form of the BK and the BFKL evolution equations. We have compared the obtained solutions to the unresummed ones as well as to the solutions provided by the {\sf BKsolver} package.
The solutions of the BK equation formulated as the integro-differential equation
and of the BK equation formulated as the double-integral equation
agree well, with the discrepancy staying below few percent.
We notice the residual dependence of the BK equation on the resummation parameter and also observe that it is less sensitive to the resummation parameter $\mu$ as compared to the BFKL equation. We attribute this phenomenon to the emergence of the saturation scale in the BK equation.
\section*{Acknowledgements}
\noindent
The  work has been partially supported by the Narodowe Centrum Bada\'n i Rozwoju with grant LIDER/02/35/L-2/10/NCBiR/2011
and the Polish National Science Centre grant DEC-2012/04/M/ST2/00240.


\end{document}